\documentclass[fleqn,twoside]{article}
\usepackage{espcrc2,amsmath}
\usepackage{graphicx}

\usepackage[figuresright]{rotating}

\newcommand{\no}{\nonumber}

\title{Tau polarization effects in the CNGS $\nu_{\tau}$ appearance
       experiments% 
       \thanks{presented by K. Mawatari in poster session 
       at the 4th International Workshop on Neutrino-Nucleus Interactions
        in the Few GeV Region (NuInt05), September 26-29, 2005, Okayama,
        Japan.}}

\author{M. Aoki\address[kek]{Theory Group, KEK, Tsukuba 305-0801,
                             Japan},
        K. Hagiwara\addressmark[kek]%
                   \address{Dept.\ of Particle and Nuclear Physics,
                     Graduate Univ.\ for Advanced Studies, Tsukuba
                     305-0801, Japan},
        K. Mawatari\addressmark[kek]%
           \thanks{present address: Korea Institute for Advanced Study,
          Seoul 130-722, Korea (kentarou@kias.re.kr)} and
        H. Yokoya\address{Dept.\ of Physics, Niigata Univ.,
                             Niigata 950-2181, Japan}
        }
       
\begin{document}

\begin{abstract}
We studied $\tau$ polarization effects on the decay distributions of
 $\tau$ produced in the CNGS $\nu_{\tau}$ appearance experiments.   
We show that energy and angular distributions for 
the decay products in the laboratory frame  
are significantly affected by the $\tau$ polarization. 
Rather strong azimuthal asymmetry about the
 $\tau$ momentum axis is predicted, which may have observable
 consequences in experiments even with small statistics.
\end{abstract}
% typeset front matter (including abstract)
\maketitle

\setcounter{footnote}{0}
%11111111111111111111111111111111111111111111111111111111111111111
\section{INTRODUCTION}

An appearance search for $\nu_{\mu}\to\nu_{\tau}$ oscillation is
expected to be performed in the CNGS (CERN Neutrino to Gran Sasso)
project~\cite{cngs}, which will start taking data from the year 2006.  
The project with ICARUS and OPERA detectors
aims to detect $\tau$-leptons produced via $\nu_{\tau}$'s which originate from 
$\nu_{\mu}$'s through oscillation, propagating from CERN to Gran Sasso in
Italy at a distance of 732~km.
ICARUS %(Imaging Cosmic and Rare Underground Signals) 
observes $\tau$
decay kinematically with liquid argon time projection chamber
\cite{icarus}, while 
OPERA %(Oscillation Project with Emulsion-tRacking Apparatus) 
does topologically with emulsion cloud chamber \cite{opera}.

Because the $\tau$ is identified by observing its decay products, and because 
the decay distribution depends crucially on its polarization,
it is important to consider the spin polarization
of $\tau$ in addition to its production cross section.

In the last report in the proceedings of NuInt04~\cite{Hagiwara:2004gs}, 
we discussed   
spin polarization of $\tau^-$ produced in neutrino-nucleon scattering at a
fixed neutrino energy. 
Here, we extend to decay distributions from produced $\tau^-$,
especially for the CNGS experiments. 
We consider $\tau^-$ production in the 
neutrino-nucleon scattering and its subsequent decays, for the decay modes 
$\tau^-\to\pi^-\nu_\tau$ and 
$\tau^-\to\ell^-\bar\nu_\ell\nu_\tau$ ($\ell=e,\,\mu$).

%22222222222222222222222222222222222222222222222222222222222222222
\section{POLARIZATION OF PRODUCED $\tau^-$}

Let us start with a brief summary of 
spin polarization of $\tau^-$ produced via neutrino: (i) at a fixed neutrino
energy~\cite{Hagiwara:2003di}; (ii) in the CNGS
experiments with their neutrino flux~\cite{Aoki:2005wb}.  

We consider $\tau^-$ production by charged current (CC) reactions off a
nucleon target: 
\begin{equation}
 \nu_\tau(k) + N(p) \to \tau^-(k') + X(p'). 
\end{equation}
For the hadronic final states $X$, we  consider three subprocesses;    
the quasi-elastic scattering (QE), the $\Delta$ resonance production 
(RES) and the deep inelastic scattering (DIS) processes.
The four-momenta are parametrized in the laboratory frame as 
\begin{align}
 k &= (E_\nu,\, 0,\, 0,\, E_\nu),\no \\
 p &= (M,\, 0,\, 0,\, 0),\\
 k'&= (E_\tau,\, p_\tau\sin\theta_\tau,\, 0,\, p_\tau\cos\theta_\tau),\no
\end{align}
and the following Lorentz invariant variables are defined
\begin{align}
 & Q^2 = -q^2 = -(k-k')^2, \quad W^2 =(p+q)^2.
\end{align}
Each subprocess is distinguished by the hadronic invariant mass $W$: 
$W=M$ for QE, $M+m_\pi <W<W_{\rm cut}$ for RES\@. 
$W_{\rm cut}$ is an artificial boundary,  
and we regard that DIS process occurs in the regions of $W>W_{\rm cut}$.
We take $W_{\rm cut}=1.4$ GeV in this report.

The differential cross section and the spin polarization vector of
produced $\tau^-$ are obtained in the laboratory frame as
\begin{align}
 &\frac{d\sigma_{\tau}}{dE_{\tau}\,d\cos\theta_{\tau}} \no \\
 &=
 \frac{G_{F}^{2}\kappa^{2}}{2\pi}
 \frac{p_{\tau}}{M}\,\big\{
 (2W_{1}+\frac{m_{\tau}^{2}}{M^{2}}\,W_{4})
 \left(E_{\tau}-p_{\tau}\cos\theta_{\tau}\right) \no \\
 &\hspace*{28pt} +W_{2}\left(E_{\tau}+p_{\tau}\cos\theta_{\tau}\right)
 -\frac{m_{\tau}^{2}}{M}\,W_{5}\no\\ 
 &\hspace*{28pt}
 +\frac{W_{3}}{M}\,(E_{\nu}E_{\tau}+p_{\tau}^{2}
 -(E_{\nu}+E_{\tau})p_{\tau}\cos\theta_{\tau})\big\}\no\\
 &\equiv \frac{G_{F}^{2}\kappa^{2}}{2\pi}
 \frac{p_{\tau}}{M}\;F,  \label{cross}
\end{align}
and
\begin{subequations}
\begin{align}
 s_{x} =&  -\frac{m_{\tau}\sin\theta_{\tau}}{2F}
 \big(2W_{1}-W_{2}+\frac{E_{\nu}}{M}\,W_{3} \no \\
 &\hspace*{60pt}
 -\frac{m_{\tau}^{2}}{M^{2}}\,W_{4}+\frac{E_{\tau}}{M}\,W_{5}\big),\\
 s_{y} =&\;0,\\
 s_{z} =&-\frac{1}{2F}\big\{
 (2W_{1}-\frac{m_{\tau}^{2}}{M^{2}}\,W_{4})
 \left(p_{\tau}-E_{\tau}\cos\theta_{\tau}\right) \no \\
 &+W_{2}\left(p_{\tau}+E_{\tau}\cos\theta_{\tau}\right)
     -\frac{m_{\tau}^{2}}{M}\,W_{5}\cos\theta_{\tau}\no\\
 &+\frac{W_{3}}{M}\,\big((E_{\nu}+E_{\tau})p_{\tau}
     -(E_{\nu}E_{\tau}+p_{\tau}^{2})\cos\theta_{\tau}\big)\big\},
\end{align} \label{spin}%
\end{subequations}%
where $G_F$ is the Fermi constant and $\kappa=M_W^2/(Q^2+M_W^2)$. 
$\vec{s}=(s_x,s_y,s_z)$ is defined in the $\tau$ rest frame in which the
$z$-axis is taken along its momentum direction and the $y$-axis is along
$\vec k \times \vec{k'}$, the normal of the scattering plane, in the
laboratory frame.  
It is  normalized as $|\vec{s}|=1/2$ for pure spin eigenstates.
$W_{i=1,\ldots,5}$ are structure functions defined 
with the generic decomposition of the hadronic tensor,
\begin{multline}
 W_{\mu\nu}(p,q)
 =-g_{\mu\nu}W_1
 +\frac{p_\mu p_\nu}{M^2}\,W_2
 -i\epsilon_{\mu\nu\alpha\beta}\frac{p^\alpha q^\beta}
 {2M^2}\,W_3  \\
  +\frac{q_\mu q_\nu}{M^2}\,W_4
  +\frac{p_\mu q_\nu+q_\mu p_\nu}{2M^2}\,W_5,
\end{multline}
where the totally anti-symmetric tensor $\epsilon_{\mu\nu\alpha\beta}$ is 
defined as $\epsilon_{0123}=1$. These functions can be estimated for
each process, QE, RES, and DIS. See explicit forms in
Ref.~\cite{Aoki:2005wb}.

\begin{figure}
\begin{center}
 \includegraphics[width=5cm,clip]{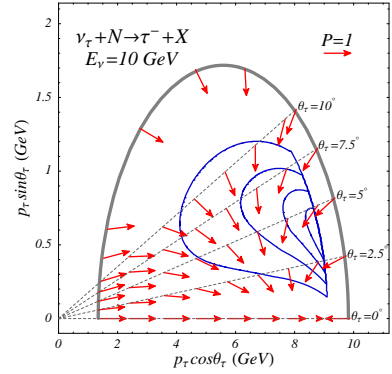}
\vspace*{-8mm}
 \caption{Contour plot of the DIS cross section for the  
$\nu_{\tau}N\to \tau^{-}X$ process at $E_{\nu}=10$ GeV in the laboratory
 frame.
The kinematical boundary is shown by a thick curve. 
The $\tau$ polarizations are shown by arrows. 
The length of the arrows gives the degree of
 polarization, and the direction of the arrows gives that of the $\tau$
spin in the $\tau$ rest frame. The size of the 100\% polarization ($P=1$) 
arrow is shown as a reference.}\label{convec}
\end{center}
\end{figure}

\noindent
{\bf (i) at a fixed neutrino energy}

In Fig.~\ref{convec}, the differential cross sections
$d\sigma_{\tau}/dp_z\,dp_T$, obtained from Eq.\ (\ref{cross}), 
at the incident neutrino energy $E_{\nu}=10$~GeV are shown as a contour map%
\footnote{It must be noted that this contour map differs from
Fig.~3 of Ref.~\cite{Hagiwara:2004gs}, where
$d\sigma_{\tau}/dE_\tau\,d\cos\theta_\tau$ is plotted.},
where $p_z=p_\tau \cos\theta_\tau$ and $p_T=p_\tau\sin\theta_\tau$.
Only the contours of the DIS cross section are plotted to avoid too much
complexity. 
Each contour gives the value of the differential cross section in
the unit of fb/GeV$^2$, e.g.,  
the innermost line is for 4 fb/GeV$^2$.
The QE process 
contributes along the boundary, and the RES contributes just
inside of the boundary. 
The contour shows that the contributions in the forward scattering angles
in the larger $p_\tau$ side are important. In that region,
the cross sections of QE and RES are also large and comparable to that
of DIS.

The polarization vector $\vec s$, Eq.\ (\ref{spin}), of $\tau^-$ is 
also shown in Fig.\ \ref{convec}.
The length of each arrow gives the degree of 
polarization ($0\leq P=2|\vec s| \leq 1$) at each kinematical point and its 
orientation gives the spin direction in the $\tau$ rest frame. 
The produced $\tau^-$'s have high degree of polarization, but their spin
directions significantly deviate from the massless limit predictions,
where all $\tau^-$'s should be purely left-handed. 
Since $s_x$ of Eq.~(\ref{spin}) turns out to be always negative, the spin
vector points to the direction of the initial neutrino momentum axis. 
Qualitative
feature of the results can be understood by considering the helicity
amplitudes in the center-of-mass (CM) frame of the scattering particles
and the effects of Lorentz boost from the CM frame to the laboratory
frame. 
See more details in Ref.~\cite{Hagiwara:2003di}.  
Let us stress that these features of the polarization of $\tau^-$ play
an important role in the following analysis.\\

\begin{figure}
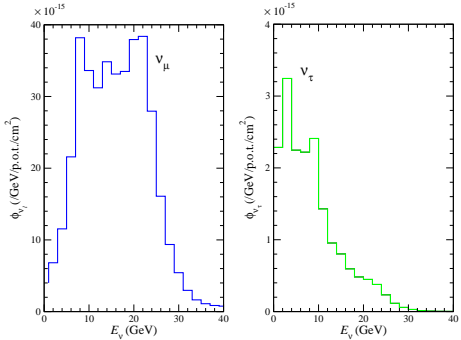

\begin{center}
 \includegraphics[height=4.5cm,clip]{fluxl}
 \includegraphics[height=4.5cm,clip]{flux_tau}
\vspace*{-8mm}
 \caption{\label{flux} Initial $\nu_{\mu}$ flux (left) and $\nu_{\tau}$
 flux followed by $\nu_{\mu}\to\nu_{\tau}$ oscillation at Gran Sasso
 (right).} 
\end{center}
\end{figure}

\noindent
{\bf (ii) in the CNGS experiments}

\begin{figure}
\begin{center}
 \includegraphics[width=5cm,clip]{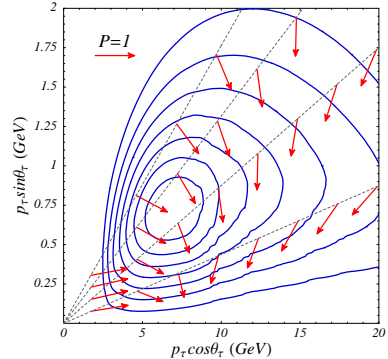}
\vspace*{-8mm}
 \caption{\label{conveccngs} Contour plot of the number of $\tau^-$
 production events in the CNGS
 experiments. The $\tau^-$ polarization are shown by the arrows.} 
\end{center}
\end{figure}

In the CNGS experiments, $\nu_\mu$ beam is produced at CERN-PS,
which is expected to deliver $4.5\times 10^{19}$ protons on target 
(p.o.t.) per year. The beam is optimized for $\nu_\tau$ appearance with
$\langle E_{\nu}\rangle=17$~GeV\@.
Figure \ref{flux} shows the expected $\nu_\tau$ flux,
\begin{equation}
 \phi_{\nu_\tau}(E_\nu)=\sum_{\ell=e,\,\mu}\phi_{\nu_\ell}(E_\nu)
 \!\times\! P_{\nu_\ell\to\nu_\tau}(E_\nu), \label{tauflux}
\end{equation}
at Gran Sasso with the baseline length of $L=732$~km from CERN\@.
Here $\phi_{\nu_\ell}(E_\nu)$ are the initial $\nu_\ell$ fluxes 
($\ell=e,\,\mu$)~\cite{cngs}, see Fig.~\ref{flux} (left),
and $P_{\nu_\ell \to\nu_\tau}(E_\nu)$ are the 
$\nu_\ell \to \nu_\tau$
oscillation probabilities in the three-neutrino model.
The fraction $\nu_e/\nu_\mu$ in the initial fluxes is less than 1\%.
In our analysis, we take the following values for the neutrino
oscillation parameters:
\begin{gather}
 \delta m^2_{12,13} = (8.2\times 10^{-5},\ 2.5\times 10^{-3})\ 
                      {\rm eV^2}, \no \\ 
 \sin^{2}{2\theta_{12,23,13}}=(0.8,\ 1,\ 0),\quad 
 \delta_{\rm MNS}=0^\circ,
\label{osc_para}
\end{gather}
with the constant matter density of $\rho=3$ g/cm$^3$.
Here we assume the so-called normal hierarchy.

Taking into account the CNGS neutrino flux shown in Fig.\ \ref{flux},
we show the distributions of events and polarization vectors of
$\tau^{-}$  on the 
$p_{\tau}\cos\theta_{\tau}$-$p_{\tau}\sin\theta_{\tau}$ plane in 
Fig.~\ref{conveccngs}. 
The initial neutrino energy is integrated out with the incoming $\nu_\tau$
flux, $\phi_{\nu_\tau}(E_\nu)$ of Eq.~(\ref{tauflux}), 
whereas it is fixed at 10 GeV in Fig.~\ref{convec}. 
The number of $\tau^-$ production events for all the QE, RES and DIS
processes is included in the contour map, where we assume 5 years 
with $4.5\times10^{19}$~p.o.t./year of the primary proton beam and
the 1.65 kton size detector, which 
are the current plan of the OPERA experiment~\cite{opera}.
Each contour gives a number of events per GeV$^2$, e.g.,  
the innermost line is for 7~events/GeV$^2$. 
The contour map shows that there are many events around $E_\tau$=10 GeV,
and around $\theta_\tau=5^\circ$. 
As for the polarization vectors, the dependence on the energy and the 
scattering angle of $\tau^-$ is rather smooth as compared to that in 
Fig.~\ref{convec} because of the integration of the incident neutrino energy. 
However, the direction of the $\tau$
polarization is still non-trivial in the region which has many events.

% 33333333333333333333333333333333333333333333333333333333333333
\section{TAU DECAY DISTRIBUTIONS}

We present our results of the decay particle
distributions from $\tau^-$ leptons produced by the CC interactions for the
CNGS experiments. 
Main feature of our analysis is to deal with the proper spin polarization
of $\tau^-$ which is calculated for each 
production phase space, shown in Fig.\ \ref{conveccngs}.
In order to show the effects of the $\tau$ polarization on the decay
distributions, we compare the results with unpolarized $\tau$ decays and
also with completely left-handed $\tau$ decays. 

The events of the decay distributions for $\rm i = \pi,\,\ell$ are given by 
\begin{align}
 \frac{dN_{\rm i}}{dE_{\rm i}d\Omega_{\rm i}}
&= A\int_{E_{\nu}^{\rm thr}}^{E_{\nu}^{\rm max}}
\!\!dE_{\nu}\,
\phi_{\nu_{\tau}}(E_{\nu}) \no\\
& \times 
\int^{1}_{\rm c_{min}}\!\!d\cos\theta_{\tau}
\int_{E_{-}}^{E_{+}}\!\!dE_{\tau}
\frac{d\sigma_\tau}{dE_{\tau}d\cos\theta_{\tau}}(E_\nu)\no \\
& \times 
\frac{1}{\Gamma_{\tau}}
\frac{d\Gamma_{\rm i}}{dE_{\rm i}d\Omega_{\rm i}}
\left(E_{\tau},\theta_{\tau},
\vec{s}(E_{\nu},E_{\tau},\theta_{\tau})\right),
\label{event}
\end{align}
where $A$ is the number of active targets, $E_{\nu}^{\rm max}$ 
is the maximum value of neutrino energy in the flux, 
$E_{\nu}^{\rm thr}=m_{\tau}+m_{\tau}^2/2M$ is the threshold energy to produce 
$\tau$ lepton off a nucleon, and the other integral ranges are given by
\begin{align}
&{\rm c_{min}}(E_{\nu})=\sqrt{1+\frac{M}{E_{\nu}}+ \frac{M^2}{E_{\nu}^2}
 - \frac{m_{\tau}^2}{4E_{\nu}^2}- \frac{M^2}{m_{\tau}^2}},\\
&E_{\pm}(E_{\nu},\theta_{\tau})=(b\pm\sqrt{b^2-ac})/a,
\end{align}
where 
\begin{align}
 a &=(E_{\nu}+M)^2-E_{\nu}^{2}\cos^{2}\theta_{\tau}, \no\\
 b &=(E_{\nu}+M)(ME_{\nu}+m_{\tau}^{2}/2), \\
 c &=
 m_{\tau}^{2}E_{\nu}^{2}\cos^{2}\theta_{\tau}+(ME_{\nu}+m_{\tau}^{2}/2)^2.\no
\end{align}
Here, $d\Gamma_{\rm i}/dE_{\rm i}d\Omega_{\rm i}$ is the decay distribution
for $\rm i = \pi,\,\ell$ in the laboratory frame.

\begin{figure}
\begin{center}
 \includegraphics[width=6.8cm,clip]{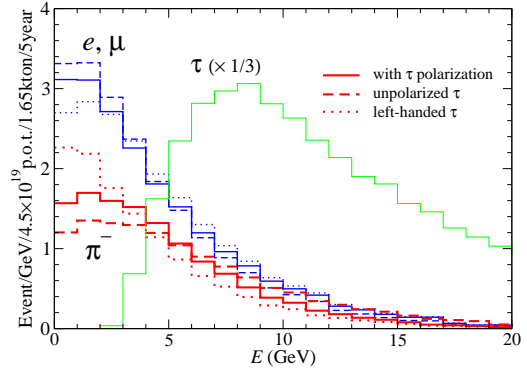}
\vspace*{-8mm}
 \caption{\label{event1} Energy distributions of $\pi^-$ (thick lines) and
 $\ell^- 
 (=e^-,\, \mu^-$) (medium-thick lines)
 in the decay of $\tau^-$ produced in the neutrino-nucleon CC interactions
 for the CNGS experiments. 
 Solid, dashed, and dotted lines shows the energy distributions
 with the predicted $\tau$ polarization, unpolarized, 
 and purely left-handed cases, respectively. The estimated number of $\tau^-$
 production is also shown by a thin solid line with respect to the
 $\tau$ energy ($E_\tau$).}
\end{center}
\end{figure}

Figure \ref{event1} shows the energy distributions of $\pi^-$ (thick
lines) and $\ell^-(= e^-,\, \mu^-)$ (medium-thick lines) decayed from
$\tau^-$ produced in the neutrino-nucleon CC interactions. 
We assume the same configuration of the experimental setup as 
Fig.~\ref{conveccngs}, i.e., 5 years running with $4.5\times10^{19}$ p.o.t.\
per year of the primary proton beam and 1.65 kton size detector for the
OPERA experiment \cite{opera}. 
For each decay mode, solid lines show the distributions from the decay
of $\tau^-$ with the predicted $\tau$ polarization. For comparison,  
dashed and dotted lines show those of unpolarized $\tau$
and purely left-handed $\tau$, respectively.
The estimated number of the $\tau^-$ production with respect to the
$\tau$ lepton energy, $E_\tau$, is also plotted as a thin solid line. 
The results are calculated by using Eq.\ (\ref{event}) 
with 100\% particle detection efficiency for simplicity.
For the above parameters, 113 events of $\tau^-$ are produced 
and 13 (20) of those decay into $\pi^-$ ($\ell^-$) mode.

The $\pi^-$ and $\ell^-$ distributions have peak in the low energy
region, and in this region the polarization dependence becomes large. 
The polarization
dependence is opposite between $\pi^-$ and $\ell^-$, and is more significant 
in $\pi^-$ than in $\ell^-$. 
In the peak region, the polarization dependence affect the distribution 
around 30\% (15\%) for the $\pi^-$ ($\ell^-$) decay mode.
Expected statistics is rather small in the current design of the
CNGS experiments. However, the likelihood probability of each event will
be affected significantly by the $\tau$ polarization effects.
The characteristic feature of our prediction is
that the produced $\tau^-$ is almost fully polarized and that it has     
non-zero transverse component of the spin vector, namely $s_x$ of 
Eq.~(\ref{spin}). 
The observed patterns of the $\pi^-$ and $\ell^-$ energy distributions
in the laboratory frame then follow from the energy angular
distributions in the polarized $\tau$ rest frame.

\begin{figure}
\begin{center}
 \includegraphics[width=6.8cm,clip]{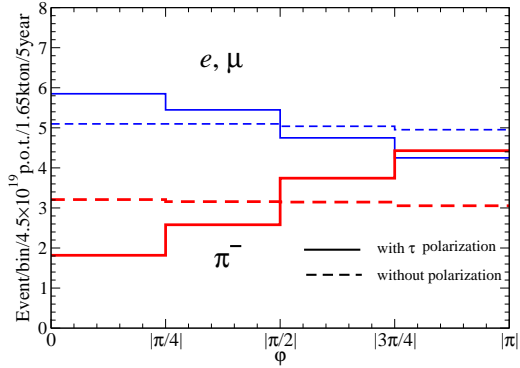}
\vspace*{-8mm}
 \caption{\label{event2}  Azimuthal angle distribution of $\pi^-$ (thick
 lines) and $\ell^-$ (thin lines). }
\end{center}
\end{figure}

Figure \ref{event2} shows the azimuthal angle distribution of $\pi^-$
(thick lines) and $\ell^-$ (thin lines). 
The azimuthal angle $\varphi_{{\rm i}=\pi,\ell}$ is given by 
$d\Omega_{\rm i}=d\cos\theta_{\rm i}\,d\varphi_{\rm i}$ in Eq.\
(\ref{event}), and is measured from the scattering plane where
$\varphi_{\rm i}=\pi/2$ is along the 
$\vec{p}_\nu \times\vec{p}_\tau$ direction in the laboratory frame,
in which the $z$-axis is taken along the direction of the $\tau$ momentum.  
Solid lines show the distributions from
$\tau^-$ with the predicted $\tau$ polarization, and dashed lines
show those from unpolarized $\tau$. The results of purely left-handed
$\tau$ are the same as those for the unpolarized $\tau$.
Since both unpolarized and purely left-handed $\tau^-$ have zero
component of perpendicular polarization, they give flat azimuthal
distributions.
The azimuthal angle distributions can be measured by tracking the
trajectory of $\tau$ leptons 
by emulsion detectors in the OPERA experiment, or by reconstructing the
hadronic cascades from neutrino-nucleon scattering. 
As is the case of energy distribution, $\pi^-$ and $\ell^-$
decay mode show the opposite feature and polarization dependence is 
clearer on $\pi^-$ mode than $\ell^-$ mode. 
At $\varphi=0$ or $|\pi|$, the dependence of the $\tau$ polarization
affects the distribution by about 47\% (16\%) for the $\pi^-$ ($\ell^-$)
decay. Even though the number of
event is limited, it may be possible to obtain a hint of such large
asymmetries.

\section{SUMMARY}

In this report, we studied the effects of the spin polarization of
$\tau^-$ produced in neutrino-nucleon scattering on the subsequent decay
distributions, for $\tau^-$ into $\pi^-$ or $\ell^-(=e^-,\,\mu^-)$ modes.
The spin polarization of $\tau^-$ produced via neutrino 
 were briefly reviewed. 
Taking into account the polarization of produced $\tau^-$, we showed
the energy and 
azimuthal angle distributions of $\pi^-$ and $\ell^-$ in the laboratory frame,
for the experimental setup of the CNGS long baseline project, 
OPERA and ICARUS experiments.

We found that the decay particle distributions in the laboratory frame 
are significantly affected by the $\tau^-$ polarization. 
Rather strong azimuthal asymmetry of $\pi^-$ and $\ell^-$ about the
$\tau$ momentum 
axis is predicted, which may have observable consequences even at 
small statistics experiments.

\section*{Acknowledgements}

K.M. would like to thank M. Sakuda for support and encouragement.

\end{document}